\documentclass[aps,pra,reprint,superscriptaddress]{revtex4-2}
\usepackage{fullpage}
\usepackage{graphicx}
\usepackage{bbm}
\usepackage{amsmath}
\usepackage{multirow}
\usepackage[utf8]{inputenc}

\begin{document}

\title{Multi-partite subspaces containing no locally inaccessible information}

\author{Sarah Croke}
\affiliation{SUPA, School of Physics and Astronomy, University of Glasgow, University Avenue, Glasgow G12 8QQ, UK}
\email{sarah.croke@glasgow.ac.uk}


\date{\today}

\newcommand{\bra}[1]{\langle #1|}
\newcommand{\ket}[1]{|#1\rangle}
\newcommand{\braket}[2]{\langle #1|#2\rangle}

\begin{abstract}
One notion of non-locality in quantum theory is the fact that information may be encoded in a composite system in such a way that it is not accessible through local measurements, even with the assistance of classical communication. Thus, contrary to the classical case, there exists information in quantum many body systems which cannot be accessed locally. We show however that, remarkably, two-dimensional subspaces do not have this property: \emph{any} physically allowed measurement on information encoded in \emph{any} two-dimensional subspace, regardless of entanglement or multi-partite structure, may be performed locally. Further, this requires only local measurement and feed-forward of classical information, readily achievable in many experimental platforms. As an application to quantum secret sharing we suggest a twist on a well known quantum information splitting protocol, which ensures that no receiving party ever has access to the full state sent, but parties must work together to perform measurements on the state. These results may have practical applications to the measurement of encoded qubits in e.g. quantum secret sharing, quantum error correction, and reveal a fundamental property unique to two-dimensional subspaces.
\end{abstract}

\maketitle

\section{Introduction}
Quantum theory exhibits at least two distinct but related notions of non-locality: entanglement, along with the related notion of Bell non-locality, which has famously played a central role in clarifying the non-local nature of quantum theory \cite{EPR,Bell,HorodeckiRMP09}; and measurement non-locality: the inaccessibility of some global information to local measurements. That these two notions of non-locality manifest in different ways was first conclusively demonstrated over twenty years ago by two rather surprising results. The first is the existence of a basis of orthogonal \emph{product} states between which local measurements cannot perfectly discriminate, even with many rounds of classical communication \cite{BennettPRA99}. This demonstrates that the absence of entanglement is not enough to ensure the local accessibility of information. The second is that \emph{any} two orthogonal states, regardless of entanglement or multi-partite structure, may be discriminated perfectly with local measurements and classical feed-forward \cite{WalgatePRL00}, showing conversely that the \emph{presence} of entanglement does not imply the local \emph{in}accessibility of information. Subsequent work shows that this latter property is peculiar to the two state case: add a third orthogonal state, and it is no longer generically true that the resulting set is perfectly locally distinguishable \cite{WalgatePRL02,HorodeckiPRL03}.

These results thus reveal something unique to two-dimensional subspaces: \cite{WalgatePRL00} tells us that any von Neumann measurement \cite{vonNeumann} (also referred to as an orthogonal, or a projective measurement \cite{NC00,Stevebook}) may be performed on information encoded in a two-dimensional subspace, regardless of the entanglement or multi-partite structure of that subspace. Von Neumann measurements however form only a subset of measurements allowed in quantum theory. In this paper we complete the picture: we show that any physically allowed measurement on a two-dimensional subspace may be implemented using only local measurement and feed-forward. It is far from obvious that this should be possible: in order to perform a generalised measurement in practice we must employ a Naimark extension \cite{NC00,Stevebook}, which involves making a von Neumann measurement on a higher dimensional space, and it is known that generically, such measurements cannot be performed locally \cite{WalgatePRL02,HorodeckiPRL03}. Nevertheless we will show by explicit construction that a local adaptive strategy always exists.

Our results illuminate a question which has seen intense study over the last couple of decades: \emph{where} is information stored in a composite quantum system, and what resources are needed to retrieve it (e.g. \cite{BennettPRA99,WalgatePRL00,WalgatePRL02,HorodeckiPRL03,PeresPRL91,GroismanJPA01,GhoshPRL01,CheflesPRA01,
VirmaniPLA01,ChenPRA01,ChenPRA02,HilleryPRA03,BadziagPRL03,FanPRL04,GhoshPRA04,JiPRA05,NathansonJMP05,
WatrousPRL05,BraunsteinPRL07,DuanPRL07,FanPRA07,CohenPRA08,BandyopadhyayJPA09,CohenPRA11,YuPRL12,PBR12,
ChildsCMP13,ChitambarPRA13,NathansonPRA13,ChitambarIEEE14,WalldenJPA14,CrokePRA2017,CrokePRA2017a,SentisPRA18,
ModiPRL18,HalderPRA18,NakahiraPRA19,HalderPRL19,WangPRA19,HaNpjqi21,BanikPRL21,SentisQuantum22,JiangPRA20,RoutPRA19})? It is known that generically \emph{some} information is always accessible locally in the subsystems: quantum information cannot be ``masked'' or hidden entirely within correlations \cite{BraunsteinPRL07,ModiPRL18}. It is also known that some information is only accessible to truly global measurements: a canonical example being the total angular momentum of a pair of spins. Our results show that, if a single qubit is encoded unitarily in a two-dimensional subspace of a multi-partite system, classical communication between subsystems is enough to ensure that \emph{all} information is accessible: in this case there is no truly non-local information, requiring global measurements to retrieve. This may have practical implications for the ease of accessing information in e.g. quantum secret sharing schemes \cite{HilleryPRA99}, and quantum error correcting codes \cite{DanielReview09}. We discuss an application to quantum secret sharing later.

Finally we note that in addition to theoretical interest, the question of which measurements may be performed locally is of considerable practical relevance. In particular, local measurements with feed-forward (i.e. with one-way classical comunication) are those measurements which can be performed without joint control or a quantum memory, and thus are readily accessible with current technology in a range of physical systems. Even in cases where there is no entanglement however, these measurements can perform significantly worse than the global optimal for certain measurement tasks \cite{CrokePRA2017,CrokePRA2017a}. It is of interest, therefore, to understand when this simpler class of measurements is sufficient for retrieval of quantum information.

We begin with some preliminaries in the following section, before introducing our main result in Section \ref{result}. We give an application to secret sharing in Section \ref{secretsharing}, and finish with discussion and conclusions in Section \ref{discussion}.

\section{Background}
Any physically allowed measurement in quantum mechanics may be described by a positive operator-valued measure (POVM) \cite{NC00,Stevebook}, that is a set of positive operators $\{ \hat{\pi}_i \geq 0 \}$ summing to the identity: $\sum_i \hat{\pi}_i = \hat{I}.$ It is a fundamental feature of quantum mechanics that measurement causes disturbance; for measurement outcome $i$, the state $\hat{\rho}$ is updated via $\hat{\rho} \rightarrow \hat{A_i} \hat{\rho} \hat{A}_i^\dagger$, where $\hat{A}_i$ is a Kraus operator \cite{Kraus} satisfying $\hat{A}_i^\dagger \hat{A}_i = \hat{\pi}_i$ \footnote{In general there may be more than one Kraus operator corresponding to each outcome, but a single Kraus operator has the property of being minimally disturbing, and is sufficient for our purposes.}. The normalisation of the resulting state gives the probability that operator $i$ was applied, and outcome $i$ registered. A set of Kraus operators $\{ \hat{A}_i \}$ thus forms an allowed operation if and only if
$\sum_i \hat{A}_i^\dagger \hat{A}_i = \hat{I}$. Note that a pure state $\ket{\psi}$ is updated to $\hat{A}_i \ket{\psi}$, given outcome $i$.

Any POVM may be decomposed as a series of two outcome operations: examples are given in \cite{AhnertPRA05,CrokeEPJD07,OreshkovPRL05}, and an explicit construction in \cite{AnderssonPRA08}. Intuitively, for an $n$ outcome measurement, in the first step we ask whether the outcome is between $0$ and $n/2-1$ or between $n/2$ and $n-1$. If this course-grained measurement is performed carefully, we can still obtain the fine-grained information, giving a particular outcome in the set, through a subsequent measurement. This is, of course, not the only way to implement a general POVM measurement, and recent work has sought to understand the resources needed to implement an arbitrary POVM \cite{OszmaniecPRL17,OszmaniecPRA19,OszmaniecNpjqi22}, however for our purposes it is sufficient that such a decomposition exists.

In order to perform any POVM on a two-level system, it is thus sufficient to be able to make transformations described by Kraus operators of the form
\begin{eqnarray}
\hat{A}_0 &=& \left( \sqrt{a} \ket{0^\prime} \bra{0} + \sqrt{b} \ket{1^\prime} \bra{1} \right) \nonumber \\
\hat{A}_1 &=& \left( \sqrt{1-a} \ket{0^{\prime \prime}} \bra{0} + \sqrt{1-b} \ket{1^{\prime \prime}} \bra{1} \right) \label{A01}
\end{eqnarray}
for any $0 \leq a,b \leq 1$ and any basis $\{ \ket{0}, \ket{1} \}$ of the two-dimensional Hilbert space. In general the bases $\{ \ket{0^\prime}, \ket{1^\prime} \}$, $\{ \ket{0^{\prime \prime}}, \ket{1^{\prime \prime}} \}$ are different from the initial basis and can be chosen freely depending on the physical implementation. 

We will also need the decomposition introduced in \cite{WalgatePRL00}: that is, for any two orthogonal bi-partite states $\ket{\chi_0}$, $\ket{\chi_1}$ there is a decomposition of the form:
\begin{eqnarray}
\ket{\chi_0} &=& \sum_i \sqrt{p_i} \ket{i}_A \ket{\eta_i}_B \nonumber \\
\ket{\chi_1} &=& \sum_i \sqrt{q_i} \ket{i}_A \ket{\eta_i^{\perp}}_B \label{walgate}
\end{eqnarray}
where $p_i, q_i \geq 0$, $\sum_i p_i = \sum_i q_i = 1$, $\{ \ket{i} \}$ are orthonormal states of system $A$, $\{ \ket{\eta_i} \}$ are arbitrary and $\braket{\eta_i}{\eta_i^\perp} = 0$. Clearly these can then be distinguished perfectly through only local measurement and feed-forward: measure basis $\ket{i}$ on system $A$, and given result $i$ make a measurement on system $B$ to distinguish perfectly between the orthogonal states $\ket{\eta_i}$, $\ket{\eta_i^\perp}$. This is readily extended to multi-partite systems, by now considering $B$ to itself be a composite system, and finding an analogous decomposition for $\ket{\eta_i}$, $\ket{\eta_i^\perp}$, once $i$ is specified.

\section{Locally Adaptive Measurement of Two-Dimensional Multi-Partite Subspaces \label{result}}
\subsection{Preliminaries: Measurement Disturbance due to Incorrect Operations}
When a measurement or operation is performed on a system, it causes disturbance, which in general is not reversible. In the scheme introduced below, we will need to understand in which cases it is still possible to perform a specified operation once a system has undergone disturbance as a result of an incorrect operation. Suppose therefore that our aim is to perform the operation given by Eq. (\ref{A01}), but instead we have performed the slightly different transformation:
\begin{eqnarray}
\hat{A}_0^\prime &=& \left( \sqrt{a^\prime} \ket{0^\prime} \bra{0} + \sqrt{b^\prime} \ket{1^\prime} \bra{1} \right), \nonumber \\
\hat{A}_1^\prime &=& \left( \sqrt{1-a^\prime} \ket{0^{\prime \prime}} \bra{0} + \sqrt{1-b^\prime} \ket{1^{\prime \prime}} \bra{1} \right). 
\end{eqnarray}
For which values of the parameters $a^\prime$, $b^\prime$ can we complete the intended operation of Eq. (\ref{A01})? If outcome ``0'' is obtained, corresponding to the application of $\hat{A}_0^\prime$, note that an operation of the form
\begin{eqnarray}
\hat{B}_{0} &=& \sqrt{\alpha_{0}} \left( \sqrt{\frac{a}{a^\prime}} \ket{0^\prime} \bra{0^\prime} + \sqrt{\frac{b}{b^\prime}} \ket{1^\prime} \bra{1^\prime} \right), \nonumber \\
\hat{B}_{1} &=& \sqrt{\alpha_{1}} \left( \sqrt{\frac{1-a}{a^\prime}} \ket{0^{\prime \prime}} \bra{0^\prime} + \sqrt{\frac{1-b}{b^\prime}} \ket{1^{\prime \prime}} \bra{1^\prime} \right), \label{krausi}
\end{eqnarray}
is sufficient to produce an effect proportional to the intended transformation $\hat{B}_i \hat{A}_0^\prime \propto \hat{A}_i$. Further, this describes an allowed operation if we can find $\alpha_{i} > 0$ such that $\hat{B}_{0}^\dagger \hat{B}_{0} + \hat{B}_{1}^\dagger \hat{B}_{1} = \hat{I}$. Defining $x = \alpha_{0} \frac{a}{a^\prime}$, $y = \alpha_{1} \frac{1-a}{a^\prime}$, we thus require $x, y \geq 0$ and:
\begin{eqnarray}
x + y &=& 1 \nonumber \\
x \frac{b}{a} + y \frac{1-b}{1-a} &=& \frac{b^\prime}{a^\prime}.
\end{eqnarray}
i.e. $b^\prime/a^\prime$ must be in the convex hull of $b/a$ and $(1-b)/(1-a)$. The requirement that we can also complete the transformation of Eq. (\ref{A01}) whenever outcome ``1'' is obtained in the first operation places a similar constraint on $\frac{1-b^\prime}{1-a^\prime}$. Supposing, without loss of generality that $a \geq b$, we therefore require
\begin{equation}
\frac{b}{a} \leq \frac{b^\prime}{a^\prime} \leq \frac{1-b}{1-a}, \qquad
\frac{b}{a} \leq \frac{1-b^\prime}{1-a^\prime} \leq \frac{1-b}{1-a}. 
\end{equation}
When both of these hold, the originally intended operation can be completed, even after a related incorrect operation has initially been performed.

This has a simple physical interpretation: note that for $a^\prime \simeq b^\prime$ the Kraus operators $A_0^\prime$, $A_1^\prime$ are approximately proportional to the identity and do not disturb the original state very much. As $a^\prime$ and $b^\prime$ become more unequal, the initial operation causes more disturbance to the initial state. The conditions above thus have the physical interpretation that the desired operation may be completed if and only if each outcome of the incorrect operation causes less disturbance than the desired operation.

\subsection{Local Adaptive Measurement Protocol for Arbitrary Two Outcome Operations}
We now proceed to our main result, in which we consider an arbitrary two-dimensional subspace of a multi-partite system, and show that all measurements on this subspace may be performed using only local measurements and feed-forward of classical outcomes. We begin our discussion with the simplest case, with just two subsystems $A$ and $B$, and in which $A$ is two-dimensional, to simplify the analysis. This is readily generalised, as we discuss later. For any basis $\ket{0_L}$, $\ket{1_L}$ of our two-dimensional subspace, there thus exists a decomposition of the form:
\begin{eqnarray}
\ket{0_L} &=& \sqrt{p} \ket{0}_A \ket{\eta_0}_B + \sqrt{1-p} \ket{1}_A \ket{\eta_1}_B, \nonumber \\
\ket{1_L} &=& \sqrt{q} \ket{0}_A \ket{\eta_0^\perp}_B + \sqrt{1-q} \ket{1}_A \ket{\eta_1^\perp}_B. \label{walgateAqubit}
\end{eqnarray}
We use the subscript $L$ to represent ``logical'' as these are not states of a single system, but encoded states spanning a multi-partite subspace. We now show how to generalize this decomposition for von Neumann measurements to perform the arbitrary two outcome operation given in Eq. (\ref{A01}), with parameters $a$, $b$, on the subspace spanned by $\{ \ket{0_L}, \ket{1_L} \}$. Thus for the remainder of this section, we suppose that the operation we intend to perform is given by:
\begin{eqnarray}
\hat{A}_0 &=& \left( \sqrt{a} \ket{0_L^\prime} \bra{0_L} + \sqrt{b} \ket{1_L^\prime} \bra{1_L} \right), \nonumber \\
\hat{A}_1 &=& \left( \sqrt{1-a} \ket{0_L^{\prime \prime}} \bra{0_L} + \sqrt{1-b} \ket{1_L^{\prime \prime}} \bra{1_L} \right), \label{A01L}
\end{eqnarray}
where $a$, $b$, $\{ \ket{0_L}, \ket{1_L} \}$ are arbitrary but given, and $\{ \ket{0_L^\prime}, \ket{1_L^\prime} \}$, $\{ \ket{0_L^{\prime \prime}}, \ket{1_L^{\prime \prime}} \}$ can be chosen in any convenient manner.

We assume without loss of generality that $a \geq b$ and $p \geq q$. To illustrate the basic idea, consider first the case in which $p=q$, and suppose we measure system $A$ in basis $\ket{0}$, $\ket{1}$: given outcome ``0'', an arbitrary state $\alpha \ket{0_L} + \beta \ket{1_L}$ is updated to:
\begin{equation}
\alpha \sqrt{p} \ket{0} \ket{\eta_0} + \beta \sqrt{p} \ket{0} \ket{\eta_0^\perp} = \sqrt{p} \ket{0} \left( \alpha \ket{\eta_0} + \beta \ket{\eta_0^\perp} \right)
\label{teleport}
\end{equation}
Similarly, for outcome ``1'', the resulting state is given by: $\sqrt{1-p} \ket{1} \left( \alpha \ket{\eta_1} + \beta \ket{\eta_1^\perp} \right)$. In each case the initial information, given by the co-efficients $\alpha$ and $\beta$, is ``teleported'' into system $B$ alone. Since it is now in system $B$, over which we have local control, any desired measurement may be performed.

At the other extreme, let us consider a case in which $p$ and $q$ are maximally different:
\begin{eqnarray}
\ket{0_L} &=& \ket{0} \ket{\eta_0} \nonumber \\
\ket{1_L} &=& \ket{1} \ket{\eta_1^\perp}.
\label{product}
\end{eqnarray}
Consider again an arbitrary initial state $\alpha \ket{0_L} + \beta \ket{1_L}$, and suppose we apply the Kraus operator $\hat{K}_0 = \sqrt{a} \ket{0}_A \bra{0}_A + \sqrt{b} \ket{1}_A \bra{1}_A$. The state of the joint system is thus updated to:
\begin{eqnarray}
&& \alpha \sqrt{a} \ket{0} \ket{\eta_0} + \beta \sqrt{b} \ket{1} \ket{\eta_1^\perp} \\
&& = \left( \sqrt{a} \ket{0_L} \bra{0_L} + \sqrt{b} \ket{1_L} \bra{1_L} \right) (\alpha \ket{0_L} + \beta \ket{1_L} ),\nonumber 
\end{eqnarray}
and the operation specified by $\hat{A}_0, \hat{A}_1$ may be achieved through acting on $A$ alone.

The general case is somewhere between these two extremes: our general strategy will be to perform some two outcome operation on system $A$, and consider the effect on the subspace of interest. We will then tailor the choice of measurement on $A$ to ensure that the intended operation $\hat{A}_0, \hat{A}_1$ can be completed, either on $A$ alone or through a subsequent measurement on $B$. Consider therefore the effect of a general operation on system $A$, described by the following pair of Kraus operators:
\begin{eqnarray}
\hat{K}_0 &=& \sqrt{c} \hat{P}_0 + \sqrt{d} \hat{P}_1, \nonumber \\
\hat{K}_1 &=& \sqrt{1-c} \hat{P}_0 + \sqrt{1-d} \hat{P}_1. \label{KA}
\end{eqnarray}
where $\hat{P}_0 = \ket{0} \bra{0}$, $\hat{P}_1= \ket{1} \bra{1}$. The effect of $\hat{K}_0$ on an arbitrary initial state is:
\begin{eqnarray}
\alpha \ket{0_L} + \beta \ket{1_L} &\rightarrow& \alpha \left( \sqrt{c} \sqrt{p} \ket{0} \ket{\eta_0} + \sqrt{d} \sqrt{1-p} \ket{1} \ket{\eta_1} \right) \nonumber \\
&& + \beta \left( \sqrt{q} \sqrt{c} \ket{0} \ket{\eta_0^\perp} + \sqrt{d} \sqrt{1-q} \ket{1} \ket{\eta_1^\perp} \right) \nonumber \\
&=& \alpha \sqrt{c p + d (1-p) } \ket{0_L^\prime} \nonumber \\
&& + \beta \sqrt{c q + d (1-q)} \ket{1_L^\prime},
\end{eqnarray}
where
\begin{eqnarray}
\ket{0_L^\prime} &=& \sqrt{\frac{c p}{c p + d (1-p)}} \ket{0} \ket{\eta_0} \nonumber \\&& + \sqrt{\frac{d (1-p)}{c p + d (1-p)}} \ket{1} \ket{\eta_1}, \nonumber \\
\ket{1_L^\prime} &=& \sqrt{\frac{c q}{c q + d (1-q)}} \ket{0} \ket{\eta_0^\perp} \nonumber \\
&&+ \sqrt{\frac{d (1-q)}{c q + d (1-q)}} \ket{1} \ket{\eta_1^\perp}, \label{primedbasis}
\end{eqnarray}
that is, the effective Kraus operator on the subspace of interest is
\begin{eqnarray}
\hat{K}_{{\rm eff},0} &=& \sqrt{c p + d (1-p) } \ket{0_L^\prime} \bra{0_L} \nonumber \\
&&+ \sqrt{c q + d (1-q) } \ket{1_L^\prime} \bra{1_L}.
\label{K0eff}
\end{eqnarray}
Now, to perform a specified two-outcome operation, given by Eq. (\ref{A01L}), clearly we would like to choose $c$, $d$ such that $a = c p + d (1-p)$, $b = c q + d (1-q)$. Solving for $c$ and $d$ in terms of $a$ and $b$ gives:
\begin{equation}
c = \frac{1-q}{p-q} a - \frac{1-p}{p-q} b, \quad d = \frac{-q}{p-q} a + \frac{p}{p-q} b.
\label{cd}
\end{equation}
This gives us an allowed operation if and only if $0 \leq c,d \leq 1$. Note that $c \geq 0$ and $d \leq 1$ hold by assumption for the chosen parameter orderings, while $c \leq 1$ and $d \geq 0$ require
\begin{eqnarray}
\frac{1-b}{1-a} &\leq& \frac{1-q}{1-p} \label{condition1} \\
\frac{b}{a} &\geq& \frac{q}{p} \label{condition2}
\end{eqnarray}
respectively. Thus if both of these hold, the desired operation can be performed on the information stored in the joint system, through operations on $A$ alone, with the above choice of parameters $c$ and $d$.

In general, of course, one or both of these may not hold, and we conclude by showing that even in this case, we can perform the desired operation, but in general we must act on system $B$ also. Suppose therefore that inequality (\ref{condition1}) does not hold, but inequality (\ref{condition2}) does hold. The above procedure leads to a value of $c$ which is greater than 1, and therefore not physically realisable. Taking $c=1$, we can nevertheless choose $d$ such that $\hat{K}_{{\rm eff}, 0} \propto \hat{A}_0$; after a little algebra we find that this corresponds to $d = \frac{-q a + p b}{(1-q) a - (1-p) b}$. It is readily verified that in this case eqn (\ref{K0eff}) becomes:
\begin{eqnarray}
\hat{K}_{{\rm eff}, 0} &=& \sqrt{\frac{p-q}{(1-q) a - (1-p) b}} \left( \sqrt{a} \ket{0_L^\prime} \bra{0_L} \right. \nonumber \\
&&  \left.+ \sqrt{b} \ket{1_L^\prime} \bra{1_L} \right).
\end{eqnarray}
Outcome ``1'' on system A is then given by $\hat{K}_1 = \sqrt{1-d} \hat{P}_1$, with an effective action on the subspace of interest given by:
\begin{eqnarray}
\hat{K}_{{\rm eff}, 1} &=& \sqrt{\frac{a-b}{(1-q) a - (1-p) b}} \left( \sqrt{1-p} \ket{0_L^{\prime \prime}} \bra{0_L} \right. \nonumber \\
&& \left. + \sqrt{1-q} \ket{1_L^{\prime \prime}} \bra{1_L} \right)
\end{eqnarray}
where $\ket{0_L^{\prime \prime}} = \ket{1}_A \ket{\eta_1}_B$, $\ket{1_L^{\prime \prime}} = \ket{1}_A \ket{\eta_1^\perp}_B$. In case of outcome ``0'' we are done, while in case of outcome ``1'', all the information is now contained in system $B$, and as we have stipulated that inequality (\ref{condition1}) does not hold, it follows that
\begin{equation}
\frac{b}{a} \leq \frac{1-q}{1-p} \leq \frac{1-b}{1-a},
\end{equation}
and the desired operation can be completed through Kraus operators such as eqn (\ref{krausi}), acting on system $B$ alone. The case in which inequality (\ref{condition1}) holds but inequality (\ref{condition2}) does not is treated analogously, choosing $d=0$, and $c$ accordingly, so that $\hat{K}_{{\rm eff},1} \propto \hat{A}_1$. Finally, if neither inequality holds, the measurement on $A$ is a projective measurement in which either outcome transforms the (partially distorted) information into system $B$, where the desired operation and any subsequent measurements may be performed through acting on $B$ alone. The strategies are summarised in Table \ref{strategies}.
\begin{table*}[]
\centering
 \begin{tabular}{||c | c | c ||} 
 \hline
 & $\frac{1-b}{1-a} \leq \frac{1-q}{1-p}$ & $\frac{1-b}{1-a} \geq \frac{1-q}{1-p}$ \\ 
 \hline\hline
 \multirow{2}{4em}{$\frac{b}{a} \leq \frac{q}{p}$} & $c= \frac{a-b}{p(1-b)-q(1-a)}, d=0$ & $c = 1, d =0$
 \\
 & Outcome 1 requires operation on $B$ & Both outcomes require operation on $B$ \\ 
 \hline
\multirow{2}{4em}{ $\frac{b}{a} \geq \frac{q}{p}$} & $c = \frac{1-q}{p-q} a - \frac{1-p}{p-q} b$, $d = \frac{-q}{p-q} a + \frac{p}{p-q} b$ & $c=1$, $d = \frac{-q a + p b}{(1-q) a - (1-p) b}$ \\
& No operation necessary on $B$ & Outcome 0 requires operation on $B$ \\
 \hline
 \end{tabular}
 \caption{Strategies for local measurement. In each case, the first step is an operation on system $A$ of the form $K_0 = \sqrt{c} \hat{P}_0 + \sqrt{d} \hat{P}_1, K_1 = \sqrt{1-c} \hat{P}_0 + \sqrt{1-d} \hat{P}_1$. The table gives the choice of parameters $c$, $d$, and the cases in which further operations on system B are required, for different relationships between the parameters $a,b,p,q$. \label{strategies}} 
\end{table*}

Returning now to the general case, where $A$ may not be two-dimensional, we use again the decomposition given in Eq (\ref{walgate}). There are several ways of constructing the desired measurement in this case: perhaps the most straight-forward is to directly generalise the discussion above. Defining $p = \sum_{i |p_i \geq q_i} p_i$, $q = \sum_{i|p_i \geq q_i} q_i$, and
\begin{equation}
\hat{P}_0 = \sum_{i |p_i \geq q_i} \ket{i} \bra{i}, \quad \hat{P}_1 = \hat{I}_A - \hat{P}_0,
\end{equation}
generically, the first step is again a measurement on $A$ of the form eqn (\ref{KA}). For parameters satisfying conditions (\ref{condition1}) and (\ref{condition2}), again this is sufficient. If these are not satisfied then we choose the parameters $c$ and $d$ as in Table \ref{strategies}. In one or both cases system $A$ is projected into the subspace with projector either $\hat{P}_0$ or $\hat{P}_1$, and is still entangled with system $B$. We simply iterate the above step in this subspace of $A$ until either the desired operation is complete, or the information is transferred (in general with some distortion) to $B$.

Finally we note that at each step, either the desired operation is completed through acting on $A$ alone or, after the operation, all the information is transferred to system $B$, which is no longer entangled with $A$. In the multi-partite case, we can consider $B$ to itself be made of two or more subsystems. It is clear then that following operations on the $i$th system, either the desired operation is completed, or the entanglement between this system and all subsequent systems is broken, and the information pushed along the chain of subsystems. Importantly, we do not need to perform any operations on the $(i+1)$th system until the $i$th system is no longer entangled with the others. This ensures that any general operation can be performed by local measurement and feed-forward, and multiple rounds of classical communication are not needed.

\subsection{Example}
To illustrate the ideas outlined above we present an example of performing a POVM measurement on information encoded in a two-dimensional subspace of a two-qubit system. We consider the BB84 measurement in which the parties wish to perform a measurement in either the eigenbasis of $\hat{\sigma}_z$ or that of $\hat{\sigma}_x$ of a qubit \cite{BB84}. We begin by constructing a decomposition of the POVM measurement into two outcome operations, and then show how to implement this in a specific encoding of a qubit into a physical two-qubit system. 

Thus, consider a single qubit with computational basis states $\ket{0}$, $\ket{1}$, and where as usual the eigenbasis of $\hat{\sigma}_x$ is given by $\ket{\pm} = \frac{1}{\sqrt2} \left( \ket{0} \pm \ket{1} \right)$. We wish to perform the measurement:
\begin{eqnarray}
\hat{\pi}_{00} &=& \frac{1}{2} \left( \ket{0} \bra{0} \right), \nonumber \\
\hat{\pi}_{01} &=& \frac{1}{2} \left( \ket{+} \bra{+} \right), \nonumber \\
\hat{\pi}_{10} &=& \frac{1}{2} \left( \ket{1} \bra{1} \right), \nonumber \\
\hat{\pi}_{11} &=& \frac{1}{2} \left( \ket{-} \bra{-} \right). \label{BB84meas}
\end{eqnarray}
In the BB84 protocol \cite{BB84}, outcomes $00$ and $01$ correspond to bit values of $0$, while $10$ and $11$ correspond to bit values of $1$. Clearly outcomes $00$ and $10$ correspond to $z$-basis measurements, while $01$ and $11$ to $x$-basis measurements. Thus the measurement returns two bits, the first of which corresponds to the bit value, the second to the basis. The sender and receiver agree on the shared bit value with certainty only if the measurement basis agrees with the basis used for encoding.

One way to perform this measurement is to choose at random between the $z$-basis and the $x$-basis, and perform a von Neumann measurement in the chosen basis. For our purposes however, to illustrate the decomposition of general POVMs into two outcome operations, we consider instead grouping outcomes as follows in the first step: $\hat{\pi}_0^\prime = \hat{\pi}_{00} + \hat{\pi}_{01}$, $\hat{\pi}_1^\prime = \hat{\pi}_{10} + \hat{\pi}_{11}$. 
The full decomposition then proceeds as follows: in the first step we perform the two outcome operation:
\begin{equation}
    \hat{A}_0 = \hat{U}_0 \hat{\pi}_0^{\prime 1/2}, \quad \hat{A}_1 = \hat{U}_1 \hat{\pi}_1^{\prime 1/2},
\label{A01BB84}
\end{equation}
where $\hat{U}_0$, $\hat{U}_1$ may be chosen in any convenient manner. Given outcome $i=0,1$ at the first step we then perform one of two possible measurements $\{ \hat{\Pi}_0^{(i)}, \hat{\Pi}_1^{(i)} \}$, in the second step, defined as:
\begin{eqnarray}
\hat{\Pi}_0^{(0)} &=& \hat{U}_0 \hat{\pi}_0^{\prime -1/2} \hat{\pi}_{00} \hat{\pi}_0^{\prime -1/2} \hat{U}_0^\dagger \nonumber \\
\hat{\Pi}_1^{(0)} &=& \hat{U}_0 \hat{\pi}_0^{\prime -1/2} \hat{\pi}_{01} \hat{\pi}_0^{\prime -1/2} \hat{U}_0^\dagger \nonumber \\
\hat{\Pi}_0^{(1)} &=& \hat{U}_1 \hat{\pi}_1^{\prime -1/2} \hat{\pi}_{10} \hat{\pi}_1^{\prime -1/2} \hat{U}_1^\dagger \nonumber \\
\hat{\Pi}_1^{(1)} &=& \hat{U}_1 \hat{\pi}_1^{\prime -1/2} \hat{\pi}_{11} \hat{\pi}_1^{\prime -1/2} \hat{U}_1^\dagger
\end{eqnarray}
Note that both $\hat{\pi}_0^\prime$ and $\hat{\pi}_1^\prime$ are full rank, so the inverse is well-defined in each case. For each $i$ the operators $\hat{\Pi}_0^{(i)}, \hat{\Pi}_1^{(i)}$ are positive and sum to the identity, by construction. The probability of obtaining outcome $ij$ (i.e. outcome $i$ in the first step, and $j$ in measurement $\{ \hat{\Pi}_j^{(i)} \}$ in the second step) on a measurement on arbitary state $\ket{\psi}$ is:
\begin{equation}
    {\rm P}(ij|\psi) = \bra{\psi} \hat{A}_i^\dagger \hat{\Pi}_j^{(i)} \hat{A}_i \ket{\psi} = \bra{\psi} \hat{\pi}_{ij} \ket{\psi},
\end{equation}
showing that this two step procedure is a valid decomposition of the measurement $\{ \hat{\pi}_{ij} \}$. This strategy corresponds to determining the bit value in the first step, and the basis in the second step.

\begin{figure}[h]
    \centering
    \includegraphics[width=60mm]{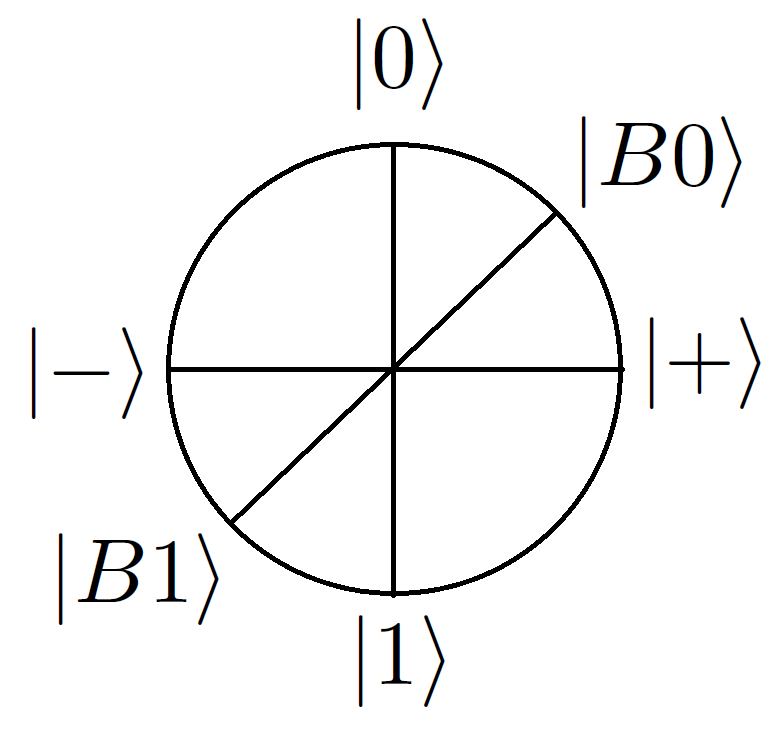}
    \caption{The $x$-$z$ plane of the Bloch sphere, showing the Breidbart basis states $\ket{B0}, \ket{B1}$, intermediate between the $x$ and $z$ bases.}
    \label{fig:Breidbart}
\end{figure}

To construct the measurements $\{ \hat{\Pi}_j^{(i)} \}$ it will be convenient to work in the eigenbasis of $\hat{\pi}_0^\prime$, $\hat{\pi}_1^\prime$, and we begin by expressing these in the Bloch sphere picture:
\begin{eqnarray}
\hat{\pi}_0^\prime &=& \frac{1}{2} \ket{0} \bra{0} + \frac{1}{2} \ket{+} \bra{+} \nonumber \\
&=& \frac{1}{4} (\hat{I} + \hat{\sigma}_z) +  \frac{1}{4} (\hat{I} + \hat{\sigma}_x) \nonumber \\
&=& \frac{1}{2} \left( \hat{I} + \frac{1}{\sqrt2} \hat{\sigma}_{\pi/4} \right),
\end{eqnarray}
where $\hat{\sigma}_{\pi/4} = \frac{1}{\sqrt2} \hat{\sigma}_z + \frac{1}{\sqrt2} \hat{\sigma}_x$. This is intermediate between $x$ and $z$, as shown in Fig. \ref{fig:Breidbart} on the Bloch sphere, with eigenvectors:
\begin{eqnarray}
\ket{B0} &=& \cos \frac{\pi}{8} \ket{0} + \sin \frac{\pi}{8} \ket{1} \nonumber \\
\ket{B1} &=& - \sin \frac{\pi}{8} \ket{0} + \cos \frac{\pi}{8} \ket{1}.
\end{eqnarray}
This basis is known as the Breidbart basis \cite{Breidbart,BB84Expt}, and is a well known attack on the BB84 protocol.
In the Breidbart basis, the BB84 states and measurements are given by:
\begin{eqnarray}
\ket{0} &=& \cos \frac{\pi}{8} \ket{B0} - \sin \frac{\pi}{8} \ket{B1}, \nonumber \\
\ket{+} &=& \cos \frac{\pi}{8} \ket{B0} + \sin \frac{\pi}{8} \ket{B1}, \nonumber \\
\ket{1} &=& \sin \frac{\pi}{8} \ket{B0} + \cos \frac{\pi}{8} \ket{B1}, \nonumber \\
\ket{-} &=& \sin \frac{\pi}{8} \ket{B0} - \cos \frac{\pi}{8} \ket{B1}. \label{BB84}
\end{eqnarray}
The eigenvalues of $\hat{\sigma}_{\pi/4}$ are $\pm 1$, thus the eigenvalues of $\hat{\pi}_0^\prime$ are $\frac{1}{2} (1 \pm \frac{1}{\sqrt2} )$. A little algebra thus shows that
\begin{eqnarray}
\hat{\pi}_0^\prime &=& \cos^2 \frac{\pi}{8} \ket{B0} \bra{B0} + \sin^2 \frac{\pi}{8} \ket{B1} \bra{B1} \nonumber \\
\hat{\pi}_1^\prime &=& \sin^2 \frac{\pi}{8} \ket{B0} \bra{B0} + \cos^2 \frac{\pi}{8} \ket{B1} \bra{B1},
\label{step1}
\end{eqnarray}
giving
\begin{eqnarray}
\hat{\pi}_0^{\prime -1/2} \left( \frac{1}{\sqrt2} \ket{0} \right) &=& \frac{1}{\sqrt2} \left( \ket{B0} - \ket{B1} \right), \nonumber \\
\hat{\pi}_0^{\prime -1/2} \left( \frac{1}{\sqrt2} \ket{+} \right) &=& \frac{1}{\sqrt2} \left( \ket{B0} + \ket{B1} \right), \nonumber \\
\hat{\pi}_1^{\prime -1/2} \left( \frac{1}{\sqrt2} \ket{1} \right) &=& \frac{1}{\sqrt2} \left( \ket{B0} + \ket{B1} \right), \nonumber \\
\hat{\pi}_1^{\prime -1/2} \left( \frac{1}{\sqrt2} \ket{-} \right) &=& \frac{1}{\sqrt2} \left( \ket{B0} - \ket{B1} \right).
\label{step2}
\end{eqnarray}
Thus the measurements in the second step $\{ \hat{\Pi}^{(i)}_j \}$, are projective measurements in the bases $\hat{U}_i \left( \frac{1}{\sqrt2} \left( \ket{B0} \pm \ket{B1} \right)\right)$.

Thus we have outlined a two-step implementation of the BB84 measurement, using operations of the form given in Eq. (\ref{A01}). We now suppose that the single qubit is encoded in a two dimensional subspace of a two-qubit system, spanned by states $\ket{0_L}$, $\ket{1_L}$ with a decomposition such as that given in Eq. (\ref{walgateAqubit}). We take this to be the Breidbart basis of the encoded qubit, and we further choose $p = 1-q = \cos^2 \frac{\pi}{8}$, so that:
\begin{eqnarray}
\ket{0_L} &=& \cos \frac{\pi}{8} \ket{0}_A \ket{\eta_0}_B + \sin \frac{\pi}{8} \ket{1}_A \ket{\eta_1}_B \nonumber \\
\ket{1_L} &=& \sin \frac{\pi}{8} \ket{0}_A \ket{\eta_0^\perp}_B + \cos \frac{\pi}{8} \ket{1}_A \ket{\eta_1^\perp}_B,
\end{eqnarray}
where $\{ \ket{\eta_0}_B, \ket{\eta_0^\perp}_B \}$, $\{ \ket{\eta_1}_B, \ket{\eta_1^\perp}_B \}$ are arbitrary but non-identical orthonormal bases of system $B$, a qubit. Note that, with this choice of $p$, $q$, a measurement in the $\{ \ket{0}, \ket{1} \}$ basis of system $A$ directly gives an implementation of $\hat{A}_{0}, \hat{A}_1$ from Eq. (\ref{A01BB84}). Explicitly, given outcome ``0", an arbitrary state $\ket{\psi}_{AB} = \alpha \ket{0_L} + \beta \ket{1}_L$ is updated to:
\begin{equation}
    \ket{0}_A (\alpha \cos \frac{\pi}{8} \ket{\eta_0}_B + \beta \sin \frac{\pi}{8} \ket{\eta_0^\perp}_B) = \hat{U}_0 \hat{\pi}_0^{1/2} \ket{\psi}_{AB}
\end{equation}
where $\hat{U}_0$ is defined such that $\hat{U}_0 \ket{0_L} = \ket{0}_A \ket{\eta_0}_B$, $\hat{U}_0 \ket{1_L} = \ket{0}_A \ket{\eta_0^\perp}_B$. Similarly it is readily seen that outcome ``1" gives an implementation of $\hat{A}_1$.

The strategy to perform the BB84 measurement on this subspace is then straight-forward: measure system A in the $\{ \ket{0}, \ket{1} \}$ basis; given outcome $i=0,1$ measure system $B$ in the $\frac{1}{\sqrt2} \left( \ket{\eta_i} \pm \ket{\eta_i^\perp} \right)$ basis. Thus, the implementation of POVM $\{ \hat{\pi}_{ij} \}$ is given by:
\begin{eqnarray}
\hat{\pi}_{00} &=& \left( \ket{0} \bra{0} \right)_A \left( \frac{1}{2} \left( \ket{\eta_0} - \ket{\eta_0^\perp} \right) \left( \bra{\eta_0} - \bra{\eta_0^\perp} \right)\right)_B \nonumber \\
\hat{\pi}_{01} &=& \left( \ket{0} \bra{0} \right)_A \left( \frac{1}{2} \left( \ket{\eta_0} + \ket{\eta_0^\perp} \right) \left( \bra{\eta_0} + \bra{\eta_0^\perp} \right)\right)_B \nonumber \\
\hat{\pi}_{10} &=& \left( \ket{1} \bra{1} \right)_A \left( \frac{1}{2} \left( \ket{\eta_1} + \ket{\eta_1^\perp} \right) \left( \bra{\eta_1} + \bra{\eta_1^\perp} \right)\right)_B \nonumber \\
\hat{\pi}_{11} &=& \left( \ket{1} \bra{1} \right)_A \left( \frac{1}{2} \left( \ket{\eta_1} - \ket{\eta_1^\perp} \right) \left( \bra{\eta_1} - \bra{\eta_1^\perp} \right)\right)_B \nonumber \\
\label{BB84implementation}
\end{eqnarray}

It is instructive to show this explicitly for one possible input state. Suppose the input state is the $\ket{0}$ state, for which the BB84 measurement, Eq. (\ref{BB84meas}), gives outcome $00$ with probability $\frac{1}{2}$, $01$ or $11$ with probability $\frac{1}{4}$ and never gives outcome $10$. According to Eq. (\ref{BB84}) this will be encoded in the two-qubit system as the state:
\begin{eqnarray}
    \ket{0} &\rightarrow& \cos \frac{\pi}{8} \ket{0_L} - \sin \frac{\pi}{8} \ket{1_L} \nonumber \\
    &=& \ket{0}_A \left( \cos^2 \frac{\pi}{8} \ket{\eta_0}_B - \sin^2 \frac{\pi}{8} \ket{\eta_0^\perp}_B \right) \nonumber \\
    && + \cos \frac{\pi}{8} \sin \frac{\pi}{8} \ket{1}_A \left(\ket{\eta_1}_B - \ket{\eta_1^\perp}_B \right).
\end{eqnarray}
For the measurement in Eq. (\ref{BB84implementation}) we obtain:
\begin{eqnarray}
{\rm P}(00) &=& \frac{1}{2} \left( \cos^2 \frac{\pi}{8} + \sin^2 \frac{\pi}{8}\right)^2 = \frac{1}{2} \nonumber \\
{\rm P}(01) &=& \frac{1}{2} \left( \cos^2 \frac{\pi}{8} - \sin^2 \frac{\pi}{8}\right)^2 = \frac{1}{2} \cos^2 \frac{\pi}{4} = \frac{1}{4} \nonumber \\
{\rm P}(10) &=& 0 \nonumber \\
{\rm P}(11) &=& \frac{1}{2} \left(2 \cos \frac{\pi}{8} \sin \frac{\pi}{8} \right)^2 = \frac{1}{2} \sin^2 \frac{\pi}{4} = \frac{1}{4}
\end{eqnarray}
as required. Note that in this implementation the measurement on the first qubit reveals the bit value, and that on the second qubit gives the basis.

Here we have carefully chosen the parameters of this example to enable a simple implementation. In a slight generalisation, suppose instead that
\begin{eqnarray}
\ket{0_L} &=& \cos \phi \ket{0}_A \ket{\eta_0}_B + \sin \phi \ket{1}_A \ket{\eta_1}_B \nonumber \\
\ket{1_L} &=& \sin \phi \ket{0}_A \ket{\eta_0^\perp}_B + \cos \phi \ket{1}_A \ket{\eta_1^\perp}_B,
\end{eqnarray}
thus $p=1-q=\cos^2 \phi$. As before we take these to be the Breidbart basis for our encoded qubit, and the first step of the measurement is given by Eq. (\ref{step1}), corresponding to a transformation of the form Eq. (\ref{A01L}) with $a=\cos^2 \frac{\pi}{8}$, $b=\sin^2 \frac{\pi}{8}$. For these parameter values we find
\begin{eqnarray}
\frac{q}{p} < \frac{b}{a} < \frac{1-b}{1-a} < \frac{1-q}{1-p}, &\;& \phi < \frac{\pi}{8} \nonumber \\
\frac{b}{a} \leq \frac{q}{p} \leq \frac{1-q}{1-p} \leq \frac{1-b}{1-a}, &\;& \phi \geq \frac{\pi}{8}.
\end{eqnarray}
Referring to Table \ref{strategies} we thus see that for $\phi < \frac{\pi}{8}$ it is always possible to choose parameters so that an operation on $A$ alone is enough to complete the first step. This will be an operation of the form Eq. (\ref{KA}), diagonal in the $\{ \ket{0}, \ket{1} \}$ basis of $A$, with the parameters $c$, $d$ chosen as in Eq. (\ref{cd}). Following this the information is still contained in a joint subspace of systems $A$ and $B$, spanned by states of the form Eq. (\ref{primedbasis}). The second step is a projective measurement, as in Eq. (\ref{step2}), and is readily implemented using the original Walgate decomposition, Eq. (\ref{walgate}). For $\phi \geq \frac{\pi}{8}$ a projective measurement on system $A$, as above, is less disturbing overall than the required transformation of the first step. Thus we simply measure in the $\{ \ket{0}, \ket{1} \}$ basis of $A$; all the information is transferred to system $B$, and the measurement can subsequently be completed through action on $B$ alone.

\section{Application: Quantum Secret Sharing \label{secretsharing}}
\subsection{Quantum information splitting: background and modified protocol}

We have shown that if quantum information is encoded in two dimensions, even if these are spread across multiple systems, any allowed measurement can be performed using only local measurements and feed-forward between parties. One implication of this is that if the state of a single qubit is split between multiple parties e.g. in a quantum secret sharing scheme, any measurement on this qubit state can be performed through the parties working together, each making separate measurements in their own labs, and communicating the results. We describe a slight twist on a well known quantum information splitting protocol in which no receiving party ever has access to the full state, but any desired measurement may be implemented by the parties working together.

In a secret sharing scheme the sender, Alice, wishes to distribute information to two or more receivers. She suspects that some of the receivers may be dishonest, but doesn't know which ones, and by forcing them to work together is assured they will behave honestly as the dishonest parties will not wish to reveal their treachery to the honest parties. Quantum secret sharing schemes use quantum mechanics to aid sharing either classical or quantum secrets. In the quantum information splitting protocol from \cite{HilleryPRA99}, a scheme to share quantum secrets, the sender, Alice, shares a GHZ state:
\begin{equation}
\ket{\rm GHZ} = \frac{1}{\sqrt{2}} \left( \ket{000} + \ket{111} \right)
\end{equation}
with Bob and Charlie, the receivers. Alice uses teleportation to transfer the state of a qubit initially in her lab to the joint subspace of Bob and Charlie's systems spanned by the states $\{ \ket{00}, \ket{11} \}$. Specifically, Alice starts with her initial state $\alpha \ket{0} + \beta \ket{1}$, and makes a Bell basis measurement on this state and her qubit of the GHZ state. Depending on the outcome of her measurement, the joint Bob-Charlie system collapses into one of the states
\begin{equation}
    \alpha \ket{00} \pm \beta \ket{11}, \qquad \beta \ket{00} \pm \alpha \ket{11}.
\end{equation}
Alice chooses one of the receivers at random, say Charlie to receive the final state, and the other party, Bob, performs a measurement in the basis $\ket{\pm} = \frac{1}{\sqrt2} \left( \ket{0} \pm \ket{1} \right)$. Charlie's system is left in one of the states:
\begin{equation}
    \alpha \ket{0} \pm \beta \ket{1}, \qquad \beta \ket{0} \pm \alpha \ket{1}.
\end{equation}
depending on both the results of Alice's and of Bob's measurement. Alice and Bob both communicate their measurement results to Charlie, enabling him to reconstruct the initial state by applying appropriate correction operators.

Security against eavesdroppers is ensured by checking some subset of states for errors, and against cheating by Bob or Charlie by Alice's ability to choose which party receives the state: if Bob intercepts both qubits and sends something else on to Charlie, he will not be detected in the instances in which Alice chooses Bob to receive the state, but will cause errors in the cases in which Alice chooses Charlie.

Note that without Alice's measurement result the joint state is maximally mixed in the $\{ \ket{00}, \ket{11} \}$ subspace, and neither Bob nor Charlie have any information about the state. If Charlie knows Alice's measurement result but not Bob's, his state is described by one of the reduced density operators:
\begin{equation}
    \rho_C = |\alpha|^2 \ket{0} \bra{0} + |\beta|^2 \ket{1} \bra{1}, \; \rho_C = |\beta|^2 \ket{0} \bra{0} + |\alpha|^2 \ket{1} \bra{1}
\end{equation}
In each case Charlie has full information about the computational basis, but the phase information is completely unknown. Only once he receives Bob's measurement result can he fully reconstruct the state. Thus the receiving parties are forced to work together to extract the quantum information. Once the protocol is complete however, the full qubit state is held by one or other of the receiving parties who can perform any desired operations without the cooperation of the other. 

We propose a slight twist on this well-established protocol, in which Alice, Bob and Charlie share the entangled state:
\begin{equation}
    \ket{\Psi}_{ABC} = \frac{1}{\sqrt{2}} \left( \ket{0}_A \ket{\chi_0}_{BC} + \ket{1}_A \ket{\chi_1}_{BC} \right)
\end{equation}
where $\ket{\chi_0}_{BC}$, $\ket{\chi_1}_{BC}$ are joint two-qubit states of Bob and Charlie's system which are orthogonal, but in general are not product states. After Alice's Bell basis measurement on her qubits, Bob and Charlie's systems are left in one of the joint states:
\begin{equation}
    \alpha \ket{\chi_0}_{BC} \pm \beta \ket{\chi_1}_{BC}, \qquad \beta \ket{\chi_0}_{BC} \pm \alpha \ket{\chi_1}_{BC},
\end{equation}
depending on the outcome of measurement. Once Alice announces her measurement result, Bob and Charlie know in which of the possibilities the information is encoded, and can perform any desired measurement through local measurements on their subsystems and feed-forward. Further, this can be achieved regardless of which party measures first, and we can assume as before that to prevent cheating Alice chooses which party measures first.  

We suggest, and prove below, two modest potential advantages of this scheme: that for a range of choices of $\ket{\chi_0}$, $\ket{\chi_1}$ there is no measurement on Bob's side that perfectly transfers the state to Charlie (or vice-versa); and that there is no basis about which either Bob or Charlie have perfect information without the co-operation of the other. Thus neither party can ever have full access to the state, and both must cooperate to extract classical information about the state.

This does not allow the parties to perform any operations which really require access to the quantum information, e.g. in which the state is subsequently used as input to a quantum register, or forwarded via teleportation to another party, but does enable any physically allowed measurement to be made on the qubit state, which need not be declared in advance. In this sense it is intermediate between quantum secret sharing protocols which share a purely classical secret and those which distribute quantum information.

\subsection{Technical details}
We will use, as before, the decomposition Eq. (\ref{walgateAqubit}) for the states of the Bob-Charlie system:
\begin{eqnarray}
\ket{\chi_0} &=& \sqrt{p} \ket{0}_B \ket{\eta_0}_C + \sqrt{1-p} \ket{1}_B \ket{\eta_1}_C, \nonumber \\
\ket{\chi_1} &=& \sqrt{q} \ket{0}_B \ket{\eta_0^{\perp}}_C + \sqrt{1-q} \ket{1}_B \ket{\eta_1^{\perp}}_C, \label{walgateQSS}
\end{eqnarray}
assuming again without loss of generality that $p \geq q$. Note that in the case $p=q$ we know that the full quantum information can be teleported from Bob to Charlie, as shown in Eq. (\ref{teleport}). We thus assume that $p \neq q$. Note also that if both states are product states, such as the case in Eq. (\ref{product}), then they are locally orthogonal for at least one party, and it follows that this party can obtain perfect information about the $\{ \ket{\chi_0}$, $\ket{\chi_1} \}$ basis, without the cooperation of the other. We thus assume that at least one of the states is entangled, which corresponds to requiring that $\ket{\eta_0} \neq \ket{\eta_1}$, and also to ruling out the case $p = 1$, $q = 0$.

We begin by showing that generically, for any choice of $\ket{\chi_0}$, $\ket{\chi_1}$ satisfying these conditions, in any joint measurement scheme Bob never perfectly transfers the state to Charlie.

\textbf{Proof:} 
An arbitrary shared state in the subspace spanned by these $\ket{\chi_0}$, $\ket{\chi_1}$ is given by
\begin{eqnarray}
\alpha \ket{\chi_0}_{BC} &+& \beta \ket{\chi_1}_{BC} \nonumber \\ &=& \ket{0}_B \left( \alpha \sqrt{p} \ket{\eta_0}_C + \beta \sqrt{q} \ket{\eta_0^\perp}_C \right) \\
&& + \ket{1}_B \left( \alpha \sqrt{1-p} \ket{\eta_1}_C + \beta \sqrt{1-q} \ket{\eta_1^{\perp}}_C \right). \nonumber \end{eqnarray}
It is convenient to re-write this as:
\begin{eqnarray}
\alpha \ket{\chi_0}_{BC} + \beta \ket{\chi_1}_{BC} &=& \ket{0}_B \left( \hat{U}_0 \hat{K}_0 \ket{\psi}_C \right) \nonumber \\
&& + \ket{1}_B \left( \hat{U}_1 \hat{K}_1 \ket{\psi}_C \right)
\label{jointBC}
\end{eqnarray}
where $\hat{U}_0$, $\hat{U}_1$, $\hat{K}_0$, $\hat{K}_1$ are operators on Charlie's system alone given by:
\begin{eqnarray}
\hat{U}_0 &=& \ket{\eta_0} \bra{0} + \ket{\eta_0^\perp} \bra{1} \\
\hat{U}_1 &=& \ket{\eta_1} \bra{0} + \ket{\eta_1^\perp} \bra{1} \\
\hat{K}_0 &=& \sqrt{p} \ket{0} \bra{0} + \sqrt{q} \ket{1} \bra{1} \\
\hat{K}_1 &=& \sqrt{1-p} \ket{0} \bra{0} + \sqrt{1-q} \ket{1} \bra{1}
\end{eqnarray}
and $\ket{\psi}$ is a state of Charlie's system alone, defined as $\ket{\psi} = \alpha \ket{0} + \beta \ket{1}$. Perfect transfer of the quantum information from the joint system to Charlie's alone is achieved if and only if the state $\ket{\psi}$ can be re-created in Charlie's system, through measurement on Bob's system. Note that neither Bob nor Charlie have performed any operations yet, Eq. (\ref{jointBC}) is merely a convenient re-writing of the state.

We can restrict to rank one measurements on Bob's system, as higher rank outcomes leave Charlie's system entangled with Bob's in general. Thus, if Bob obtains a measurement outcome proportional to a projector onto the state $\ket{\phi}$, then Charlie's system is left in the state
\begin{equation}
    \left(\braket{\phi}{0} \hat{U}_0 \hat{K}_0 + \braket{\phi}{1} \hat{U}_1 \hat{K}_1 \right) \ket{\psi}.
\end{equation}
Defining $\hat{K}_\phi = \braket{\phi}{0} \hat{U}_0 \hat{K}_0 + \braket{\phi}{1} \hat{U}_1 \hat{K}_1$, Charlie's (non-normalized) final state is given by $\ket{\psi_f} = \hat{K}_\phi \ket{\psi}$.

Noting that Charlie may perform a unitary correction operation, this procedure successfully transfers the state to Charlie's system alone if and only if $\hat{K}_\phi$ is proportional to a unitary operator, i.e. if $\hat{K}_\phi^\dagger \hat{K}_\phi \propto \hat{I}$. Noting that $\hat{K}_0$, $\hat{K}_1$ are Hermitian, we find:
\begin{eqnarray}
\hat{K}_\phi^\dagger \hat{K}_\phi &=& |\braket{\phi}{0}|^2 \hat{K}_0^2 + \braket{0}{\phi} \braket{\phi}{1} \hat{K}_0 \hat{U}_0^\dagger \hat{U}_1 \hat{K}_1 \nonumber \\
&& + \braket{1}{\phi} \braket{\phi}{0} \hat{K}_1 \hat{U}_1^\dagger \hat{U}_0 \hat{K}_0 + |\braket{\phi}{1}|^2 \hat{K}_1^2 \label{Kphi}
\end{eqnarray}
Choosing the basis in which $\hat{K}_0$, $\hat{K}_1$ are diagonal, we require:
\begin{eqnarray}
\bra{0} \hat{K}_\phi^\dagger \hat{K}_\phi \ket{0} &=& \bra{1} \hat{K}_\phi^\dagger \hat{K}_\phi \ket{1}, \nonumber \\
\bra{0} \hat{K}_\phi^\dagger \hat{K}_\phi \ket{1} &=& \bra{1} \hat{K}_\phi^\dagger \hat{K}_\phi \ket{0} = 0.
\end{eqnarray}

Considering first the off-diagonal elements, simplifying the latter condition gives:
\begin{eqnarray}
&&\braket{0}{\phi} \braket{\phi}{1} \sqrt{p} \sqrt{1-q} \braket{\eta_0}{\eta_1^\perp} \nonumber \\
&&+ \braket{1}{\phi} \braket{\phi}{0} \sqrt{1-p} \sqrt{q} \braket{\eta_1}{\eta_0^\perp} = 0
\end{eqnarray}
Note that in two dimensions $|\braket{\eta_0}{\eta_1^\perp}|=|\braket{\eta_1}{\eta_0^\perp}|$, and we can always choose $\ket{\phi}$ such that
\begin{equation}
\braket{0}{\phi} \braket{\phi}{1}\braket{\eta_0} {\eta_1^\perp} = -\braket{1}{\phi} \braket{\phi}{0}\braket{\eta_1} {\eta_0^\perp}
\end{equation}
Thus the requirement becomes:
\begin{equation}
\braket{0}{\phi} \braket{\phi}{1}\braket{\eta_0} {\eta_1^\perp} \left(\sqrt{p} \sqrt{1-q} - \sqrt{q} \sqrt{1-p} \right) = 0
\end{equation}
This is satisfied if any of the following hold: $\braket{0}{\phi} = 0$, $\braket{1}{\phi} = 0$, $\braket{\eta_0}{\eta_1^\perp} = 0$, or $p=q$. Note that $\braket{\eta_0}{\eta_1^\perp} = 0$ implies $\ket{\eta_0} = \ket{\eta_1}$. As we have ruled out this case, along with the case $p=q$, the only possibilities are $\braket{0}{\phi} = 0$, or $\braket{1}{\phi} = 0$. From Eq. (\ref{Kphi}) these cases correspond respectively to $\hat{K}_\phi^\dagger \hat{K}_\phi$ equal to $\hat{K}_0^2$ or $\hat{K}_1^2$, neither of which is proportional to the identity for $p \neq q$. Thus we conclude that the quantum information cannot be perfectly transferred to Charlie through a measurement on Bob's system.

The second advantage, which we prove next, is that Charlie does not have complete information about \emph{any} basis without Bob's cooperation. 

\textbf{Proof:} We again use Eq. (\ref{jointBC}), from which we find that the reduced density operator for $C$ alone is given by:
\begin{equation}
    \hat{\rho}_C = \hat{U}_0 \hat{K}_0 \ket{\psi} \bra{\psi} \hat{K}_0 \hat{U}_0^\dagger + \hat{U}_1 \hat{K}_1 \ket{\psi} \bra{\psi} \hat{K}_1 \hat{U}_1^\dagger.
\end{equation}
Charlie can apply an arbitrary unitary correction operator, which we denote $\hat{V}$, and complete information about a basis $\{ \ket{\phi}, \ket{\phi^\perp} \}$ is therefore available to Charlie without the cooperation of Bob if and only if there is a choice of $\hat{V}$ such that
\begin{eqnarray}
{\rm P}(\phi) &=& \bra{\phi} \hat{V} \hat{\rho}_C \hat{V}^\dagger \ket{\phi} = |\braket{\phi}{\psi}|^2, \nonumber \\
{\rm P}(\phi^\perp) &=& \bra{\phi^\perp} \hat{V} \hat{\rho}_C \hat{V}^\dagger \ket{\phi^\perp} = |\braket{\phi^\perp}{\psi}|^2.
\end{eqnarray}
It is convenient to re-write the condition for $\ket{\phi}$ as ${\rm P}(\phi) = \bra{\psi} \mathcal{E} \left( \ket{\phi} \bra{\phi} \right) \ket{\psi}$ where:
\begin{eqnarray}
\mathcal{E}(\ket{\phi} \bra{\phi}) &=& \hat{K}_0 \hat{U}_0^\dagger \hat{V}^\dagger \ket{\phi} \bra{\phi} \hat{V} \hat{U}_0 \hat{K}_0 \nonumber \\
&& + \hat{K}_1 \hat{U}_1^\dagger \hat{V}^\dagger \ket{\phi} \bra{\phi} \hat{V} \hat{U}_1 \hat{K}_1.
\end{eqnarray}
It follows that ${\rm P}(\phi)$ is preserved for all initial states $\ket{\psi}$ if and only if $\mathcal{E} \left( \ket{\phi} \bra{\phi} \right) = \ket{\phi} \bra{\phi}$, which in turn requires
\begin{equation}
\hat{K}_0 \hat{U}_0^\dagger \hat{V}^\dagger \ket{\phi} \propto
    \hat{K}_1 \hat{U}_1^\dagger \hat{V}^\dagger \ket{\phi} \propto \ket{\phi},
\end{equation}
with an equivalent requirement for $\ket{\phi^\perp}$. Noting that we have previously ruled out the case $p=1$, $q=0$, it follows that at least one of $\hat{K}_0$ or $\hat{K}_1$ is full rank. Let us suppose $\hat{K}_1$ is full rank, and therefore invertible. Applying $\hat{U}_1 \hat{K}_1^{-1}$ we require:
\begin{eqnarray}
\hat{U}_1 \hat{K}_1^{-1} \hat{K}_0 \hat{U}_0^\dagger \left(\hat{V}^\dagger \ket{\phi} \right) & \propto & \hat{V}^\dagger \ket{\phi} \nonumber \\
\hat{U}_1 \hat{K}_1^{-1} \hat{K}_0 \hat{U}_0^\dagger \left(\hat{V}^\dagger \ket{\phi^\perp} \right) & \propto & \hat{V}^\dagger \ket{\phi^\perp}.
\end{eqnarray}
In other words both $\hat{V}^\dagger \ket{\phi}$ and $\hat{V}^\dagger \ket{\phi^\perp}$ are eigenvectors of the operator $\hat{N} = \hat{U}_1 \hat{K}_1^{-1} \hat{K}_0 \hat{U}_0^\dagger$. As these are orthonormal, this is only possible if $\hat{N}$ is a normal operator, i.e. commutes with its Hermitian conjugate:
\begin{eqnarray}
[ \hat{N}, \hat{N}^\dagger ] &=& \hat{U}_1 \hat{K}_1^{-1} \hat{K}_0^2 \hat{K}_1^{-1} \hat{U}_1^\dagger -\hat{U}_0 \hat{K}_0 \hat{K}_1^{-2} \hat{K}_0 \hat{U}_0^\dagger \nonumber \\
&=& \left(\frac{p}{1-p} \ket{\eta_1} \bra{\eta_1} + \frac{q}{1-q} \ket{\eta_1^\perp} \bra{\eta_1^\perp} \right) \nonumber \\
&& -\left( \frac{p}{1-p} \ket{\eta_0} \bra{\eta_0} + \frac{q}{1-q} \ket{\eta_0^\perp} \bra{\eta_0^\perp} \right) \nonumber \\
&=& 0.
\end{eqnarray}
This can only hold if $\ket{\eta_0} = \ket{\eta_1}$; or if both $\ket{\eta_0} = \ket{\eta_1^\perp}$ and $p=q$. Both these cases are excluded, and we thus conclude that for a generic choice of $\ket{\chi_0}$, $\ket{\chi_1}$, excluding the cases outlined previously, there is no basis about which Charlie has perfect information without Bob's cooperation.

Note that there will always be some measurements that can be completed by one party alone; if the measurement is weak enough, as we have seen earlier, it can be completed by the first party alone. Nonetheless, for any strong (rank one) measurement the parties are forced to co-operate, and the two parties obtain complementary information. The amount of information obtained by each party depends on the particular measurement and the chosen basis for encoding $\ket{\chi_0}$, $\ket{\chi_1}$. Optimal choices of $\ket{\chi_0}$, $\ket{\chi_1}$, e.g. to ensure the information is shared as evenly as possible are left for future study.

\section{Discussion \label{discussion}}
In this paper we have shown that any measurement on a two-dimensional subspace can be performed using the simplest possible local measurement scheme, requiring only local measurements and classical feed-forward. This has practical significance as such measurements, requiring neither joint operations nor quantum memory, can be readily performed with current technologies in a range of physical systems. This is a generic feature of two-dimensional subspaces only, but it would be interesting to find non-trivial examples of completely locally measurable subspaces of dimension higher than two. We have also given an application to secret sharing, in which the qubit is never transferred to just one party, but measurements are performed in cooperation, and some information about the measurement result is available to each party.

Few general statements can be made about tasks that are possible with only local measurements, despite the fact that this question has seen much study \cite{BennettPRA99,WalgatePRL00,WalgatePRL02,HorodeckiPRL03,PeresPRL91,GroismanJPA01,GhoshPRL01,CheflesPRA01,
VirmaniPLA01,ChenPRA01,ChenPRA02,HilleryPRA03,BadziagPRL03,FanPRL04,GhoshPRA04,JiPRA05,NathansonJMP05,
WatrousPRL05,BraunsteinPRL07,DuanPRL07,FanPRA07,CohenPRA08,BandyopadhyayJPA09,CohenPRA11,YuPRL12,PBR12,
ChildsCMP13,ChitambarPRA13,NathansonPRA13,ChitambarIEEE14,WalldenJPA14,CrokePRA2017,CrokePRA2017a,SentisPRA18,
ModiPRL18,HalderPRA18,NakahiraPRA19,HalderPRL19,WangPRA19,HaNpjqi21,BanikPRL21,SentisQuantum22,JiangPRA20,RoutPRA19}. It is known that optimal discrimination, or hypothesis testing, of any two pure states may be performed locally \cite{WalgatePRL00,VirmaniPLA01,ChenPRA01,ChenPRA02,JiPRA05}. For more than two hypotheses, there are limited examples for which local measurements can achieve optimal performance \cite{CheflesPRA01,NathansonPRA13,ChitambarIEEE14,WalldenJPA14,SentisPRA18,NakahiraPRA19,SentisQuantum22}. Our work implies that as long as the states span only two dimensions, optimal performance may be achieved locally for any number of hypotheses. It also proves that there can be no non-locality in the discrimination of two pure states, according to any conceivable figure of merit, which was left as an open question in \cite{JiPRA05}.

There exist subspaces of composite quantum systems containing no product states \cite{BennettPRL99,Parthasarathy04,BhatIJQI06,WalgateJPA08}, so-called completely entangled subspaces. It is interesting to note that since all two dimensional subspaces are completely locally measurable, and subspaces of sufficiently small dimension are generically completely entangled \cite{WalgateJPA08}, almost all two-dimensional subspaces have both properties. Thus two-dimensional subspaces seem to be rather special: they are forbidden from displaying one facet of quantum non-locality, while maximally embodying the other.

\acknowledgements{This work was supported by a Leverhulme fellowship (RF-2020-397). This research was supported in part by Perimeter Institute for Theoretical Physics. Research at Perimeter Institute is supported by the Government of Canada through the Department of Innovation, Science and Economic Development and by the Province of Ontario through the Ministry of Colleges and Universitites. I am grateful to Robin Blume-Kohout for fruitful initial discussions.}

\end{document}